\documentclass[12pt, preprint]{aastex}
\usepackage{graphicx}

\shorttitle{CME-streamer interactions and type II radio burst sources}

\shortauthors{Feng et al.}

\usepackage{color}

\begin{document}
%\begin{CJK*}{GBK}{song}

\title{Radio signatures of CME-streamer interaction and
source diagnostics of type II radio burst}

\author{S. W. Feng\altaffilmark{1,2,3}, Y. Chen\altaffilmark{1},
X. L. Kong\altaffilmark{1}, G. Li\altaffilmark{4,1}, H. Q. Song
\altaffilmark{1}, X. S. Feng\altaffilmark{2}, and Ying
Liu\altaffilmark{5}} \altaffiltext{1}{Shandong Provincial Key
Laboratory of Optical Astronomy and Solar-Terrestrial Environment,
School of Space Science and Physics, Shandong University at
Weihai, Weihai 264209, China} \altaffiltext{2}{SIGMA Weather
Group, State Key laboratory for Space Weather, Center for Space
Science and Applied Research, Chinese Academy of Sciences, Beijing
100190, China} \altaffiltext{3}{College of Earth Sciences,
Graduate School of Chinese Academy of Sciences, Beijing 100049,
China} \altaffiltext{4}{Department of Physics and CSPAR,
University of Alabama in Huntsville} \altaffiltext{5}{Space
Sciences Laboratory, University of California, Berkeley, CA 94720,
USA}

\begin{abstract}
It has been suggested that type II radio bursts are due to
energetic electrons \textbf{accelerated} at coronal shocks. Radio observations,
however, have poor or no spatial resolutions to pinpoint the exact
acceleration locations of these electrons. In this paper, we
discuss a promising approach to infer the electron acceleration
location by combining radio and white light observations.
The key assumption is to relate specific morphological
features (e.g. spectral bumps) of the dynamic spectra of type II
radio bursts, to imaging features (e.g. CME going into a
streamer) along the CME (and its driven shock) propagation.
In this study, we examine the CME-streamer interaction
for the solar eruption dated on 2003 November 1.
The presence of spectral bump in the relevant type II radio
burst is identified, which is interpreted as a natural result of
the shock-radio emitting region entering the dense streamer
structure. The study is useful for further determinations of the
location of type II radio burst and the associated
electron acceleration by CME-driven shock.
\end{abstract}


\section{Introduction}

Type II solar radio bursts are narrow stripes present in the
metric to kilo-metric wavelength range drifting gradually from
higher to lower frequencies as revealed from the solar radio
dynamic spectra (Wild 1950; Wild et al., 1954; Nelson {\&}
Melrose, 1985). It is generally believed that these bursts are
excited by energetic electrons accelerated at magnetohydrodynamic (MHD) shocks driven by solar
eruptions via the plasma emission process (Ginzburg {\&}
Zheleznyakov, 1958). The emitting frequencies are close to the
local plasma frequency and/or its harmonic. There exists a
large body of works in the literature studying this type of solar
radio bursts (see, e.g., Dulk, 1985; Pick {\&} Vilmer, 2008, and
references therein). Nevertheless, the generation mechanism and the
source location of coronal metric type II radio bursts remain as a
fiercely debated problem. The focus of the debate is whether coronal
metric type II bursts are associated with blast waves driven by the
flare heating process, or with a piston-driven shock at the CME nose front
and/or flank (e.g., Cliver et al., 1999; Oh et al., 2007; Vr\v snak {\&} Cliver, 2008;
Magdaleni\' c et al., 2008; Liu et al., 2009). At present, the most direct
 method to resolve this issue is via radio imaging observations with
 radioheliographs, like the Teepee Tee Array of the Clark Lake Radio
Observatory, the Culgoora Radioheliograph, and the Nancay Radioheliograph
(e.g., Gergely et al., 1983;  Maia et al., 2000;
for reviews see Pick, 1999; Pick {\&} Vilmer, 2008).
However, simultaneous observations of white light/EUV (extreme ultraviolet) imagings of CME shocks
 and metric type II radio imagings have been rare.

A promising approach to resolve the above issue is to establish
physical connections between the spectral shape from the radio
data and certain eruptive processes observed using solar imaging
instruments in the white light and EUV wavelengths. Since the
type II emission frequency is largely determined by the
coronal electron density along the shock path,
any coronal density variation along the path will affect the shape of
the radio emission in the corresponding
dynamic spectrum. It is well known that
coronal streamers are the brightest dense structure extending all
the way from the solar surface to interplanetary space, frequently
interacting with the CME ejecta and related disturbances (e.g.,
Subramanina et al.,1999; Sheeley et al., 2000; Wang {\&} Sheeley,
2007; Bemporad et al., 2008, 2010; Chen et al., 2010; Feng et al.,
2011). Previous studies on radio bursts (e.g., Reiner et al.,
2003; Cho et al., 2007, 2008, 2011) have illustrated the importance
of CME-streamer interaction region as a source for type II bursts.

In a recent study, Kong et al. (2012) reported an
interesting type II event occurred on 2011 March 27. In that event
the type II radio emission showed a ``sudden'' transition from a
relatively slow drift to a much fast drift. \textbf{By analyzing
simultaneous EUV and coronagraph imaging observations, Kong et al. (2012)
inferred that the observed sudden spectral transition was a result
of the transit of an eruption-driven shock across the streamer boundary, where the
density drops sharply, from inside of the streamer.} In this study we examine another type II
event which shows a ``bump-like'' feature instead of
a ``break''.  We argue that such a feature is a result of a CME-driven shock
crossing the dense streamer from outside.

According to previous observational and numerical studies, the
streamer can be several times denser than surrounding solar wind
environment (e.g., Habbal et al., 1997; Parenti et al., 2000;
Strachan et al., 2002; Chen et al., 2001; Li et al., 2006).
Therefore, the passage of a shock crossing a nearby streamer can
result in a noticeable effect on the shape of the type II emission in the
dynamic spectrum. Generally speaking, the type II strips may
deviate away from their preceding declining trend and get elevated
temporarily depending on the shock speed, the transit direction,
and the exact density gradient along the shock path.

It is the aim of this study to identify a type II radio burst that
bears the afore-mentioned evolutionary feature in the dynamic spectrum,
and to establish the physical connection of the feature to
eruptive processes recorded by coronagraphs, and to further infer
the electron accelerating site responsible for the radio burst.

The paper is organized as follows. In Section 2, we briefly
discuss the solar radio spectrographs and coronagraphs used in
the study. In Section 3 we present the type II
event, determine its relationship with the associated mass
eruption, estimate the radio source, and finally consider other
radio features  possibly relevant to the CME-driven shock-streamer interaction.
A summary with some brief discussion is
provided in Section 4.

\section{Observational data}

The radio data used in this study are from WIND/WAVES (Bougeret et
al., 1995), the Bruny Island Radio Spectrograph (BIRS;
Erickson, 1997), and the Learmonth (LEAR; Kennewell \& Steward, 2003).
The ranges of frequency coverage and temporal
resolution are listed in the second to fourth columns of Table 1.
The corona imaging data are from LASCO (Large Angle
Spectrographic Coronagraph) C2 on board the SOHO (Solar and
Heliospheric Observatory) Spacecraft (Brueckner et al., 1995) and the
Mark-IV (MK4) coronagraph operated by Mauna Loa Solar Observatory
(Elmore et al., 2003). Their fields of view and
observational cadences are given in the last two columns of
Table 1. Previous radio imaging studies have shown
that the metric type II bursts can start as low as a few tenths
of solar radii above the photosphere (e.g. Pohjolainen et al. 2008;
 Magdaleni\'c et al. 2008; Nindos et al. 2011), so we focus mainly
 on the MK4 observation in our study.

%________________________________________ Table 1
\begin{table*}\centering
\begin{minipage}{140mm}
\caption{Parameters of the instruments used in this
study.  Ranges of frequency coverage and temporal resolution of
the radio spectrographs are listed in the second to fourth
columns, fields of view and temporal cadences of the coronagraphs
are listed in the last two columns.} \label{Tab:table1}
\begin{center}
\begin{tabular}{lcccccc}
\hline\noalign{\smallskip} \hline\noalign{\smallskip}
Instruments &\multicolumn{3}{c}{Radio spectrograph}& \multicolumn{2}{c}{Coronagraph}\\
 \cline{2-4}  \cline{6-7}
            &  WAVES &BIRS    &LEAR & & MK4   &LASCO    \\
            &  RAD2  &        &      & &      &C2       \\

\hline\noalign{\smallskip}

Observational    &1.07-13.8  &5-62.5  &25-180&& 1.14-2.85 &2.1-6.0 \\
range            &  (MHz)  &(MHz)   & (MHz)     && (R$_\odot$)  &   (R$_\odot$)  \\
\hline\noalign{\smallskip}
Temporal         &60 &3  & 3&& 3&12-36 \\
resolution       &  (sec) &(sec)&(sec)   &&(min) & (min)      \\
\noalign{\smallskip}\hline
\end{tabular}\end{center}
\end{minipage}
\end{table*}
%________________________________________ Table 1

\section{Radio burst and CME-streamer interaction on 2003 November 1}

On 2003 November 1, an M3.2 solar flare from NOAA AR10486 erupted
from S12W60 in heliographic coordinates. The flare eruption was
between 22:26 UT - 22:49 UT with the GOES X-ray flux peaked at
22:36 UT. The associated CME was first observed by MK4 and LASCO
C2 at 22:33:06 UT and 23:06:53 UT as shown in Figures 1(a) and 1(b). The
arrows indicate the location of the CME leading edges which were ~1.3
R$_\odot$ and 3.8 R$_\odot$.  Their CPAs (central position angles)
were 260$^\circ$ and 245$^\circ$, respectively. From these data we
can deduce that the CME nose propagated slightly non-radially
with a linear speed of 878 km s$^{-1}$. According to the online
CDAW (Coordinated Data Analysis Workshops) catalog (Yashiro et al., 2004),
the CME linear speed was 899 km s$^{-1}$, consistent with the above estimate.
Figure 1(c) presents the MK4 data at 22:45:50 UT. We can see that the
CME erupted from within a multi-streamer system with a bright
arc-like leading edge, which expanded rapidly from 22:33:06 UT to 22:44:50 UT.
During the expansion, the CME pushed the surrounding streamers
aside.
%We will continue to discuss the details of the
%CME-streamer interaction as we proceed.
Figure 1(d) is the difference image obtained from the LASCO C2
observations at 23:06:53 UT and 22:30:05 UT. According to this
image and the one obtained later a diffusive structure in front
 of the bright CME ejecta can be clearly identified.
% which resulted in apparent deviations of nearby coronal rays.
We believe that this structure corresponds to the
shock driven by the eruption (c.f., Vourlidas et al., 2003).
\textbf{There were two preceding CMEs, first observed in C2 field of view at
14:54:05 UT and  21:30:08 UT with CPAs being about
274$^\circ$ and 318$^\circ$, respectively.} We discuss
the connection between these two events and our event at the start of Sec. 3.2.

\subsection{Spectral bump of the type II radio burst}

In Figure 2, we show the radio dynamic spectrum observed during
the \textbf{above-mentioned solar eruption}. The spectrum is given by combining the
data from all three radio spectrographs listed in Table 1. There
was an obvious type III burst at 22:33 UT, followed by two strong
stripes of type II emission corresponding to the fundamental (F)
and harmonic (H) branches. The type II lasted for about 28 minutes
extending from $\sim$ 140 MHz to 10 MHz. The average frequency
drift rate of the F branch was $\sim$ -0.08 MHz s$^{-1}$. In the
paper of Cane {\&} Erickson (2005), this burst was taken as a typical metric
type II event with a relatively large and continuous frequency
coverage. In this study, we focus on the radio features observed
between 22:44 UT and 22:54 UT.

At 22:44 UT, both F and H branches started to be intermittent and
 \textbf{this intermittence lasted till} 22:48 UT, when the emission
 became continuous again with a clear
band splitting for the H branch. Band splitting phenomenon has
 been discussed, e.g., by Smerd (1974) and Vrs\v nak et al. (2001).
In the mean time, a large change of the spectral slope was clearly
seen at about $\sim$ 40 MHz of the H branch.  The average frequency
drift of this branch, $\sim -0.4$ MHz s$^{-1}$ between 22:34 UT and 22:44 UT,
decreased to $ \sim -0.04$ MHz s$^{-1}$ during the spectral
plateau from 22:44 UT to 22:52 UT. After the plateau, the
magnitude of the average drift rate increased slightly to
$\sim$ 0.06 MHz s$^{-1}$. The two solid-dashed lines in the figure are fittings
to the temporal evolution of the F and H branches before 22:44 UT using the
one-fold Saito density model (Saito, 1970) with a radial
shock propagation speed of 900 km s$^{-1}$. The spectral plateau
lies clearly above the black curves, which we define as the type
II spectral bump. It is the main focus of this study.

\subsection{CME-streamer interaction and the type II spectral bump}

Since the emission frequency of type II bursts depends on the plasma density along the
shock path, the most straightforward explanation of the spectral bump is that the shock
was passing through a high density coronal structure.
It is possible that such dense structure could be coronal disturbances
caused by preceding eruptions. Indeed, as mentioned before, there existed two preceding
CMEs to our CME. The first CME took place $\sim$ 8 hours before our CME. That CME is possibly
too early to affect the shape of the studied type II radio emission.
This is confirmed by the online MK4 animations for both the white-light
and difference data in the time period of 17:22:29 UT to 22:00:37 UT.
We can see that there were no observable signatures of this CME in the
MK4 FOV from the beginning of the animations. The second
event did not cause strong disturbances to the equatorial streamer
as a result of its source location being outside of the streamer and its limited
angular span. The CME left the MK4 FOV $\sim$ 30 minutes before the
 start of the radio bursts. Therefore, it is unlikely that the high density
structure accounting for the observed spectral bump was directly associated
with the disturbances of these two earlier eruptions.

According to the coronagraph observation, the CME collided with nearby streamers
from both flanks. \textbf{Therefore, interaction of shock wave with both the
northern and southern streamers can cause the spectral bump. To further discern the
interaction with which streamer is responsible for the type II spectral bump,}
in Figure 3 we present ten base-difference images
(coronal images subtracted the pre-eruption corona image at 22:30:06 UT)
from MK4 observation for the period of 22:33:06 UT to 23:59:35 UT
with the cadence of 3 minutes.

We first analyze the details of the CME (shock) interaction with both streamers.
\textbf{As seen from Figure 3 and the online animations, both the spatial
location of streamers and the CME-propagation seem to be asymmetric. The shock wave
first encountered northern streamer and a few minutes later also southern streamer.}
Black arrows in the figure indicate the locations where the
 \textbf{streamer deflections were first observed}. The deflections were present even in the last
 panel of this figure. As measured from Figure 3(c), the stand-off distance
 of the shock along the northern streamer is about
 0.3 R$_\odot$. From Figure 3(d), the southern streamer seemed to be first
 deflected by the CME shock wave at 22:41:54 UT. The deflection was very weak
 yet noticeable\textbf{ (see online animation) and it can be} considered
 as the first signature of the interaction between the shock and the southern streamer.
  Note that the spectral bump started around 22:46 UT. The lack of the change
  in the radio emission during the interaction of the shock and the northern
  streamer suggests that the type II radio emission came from the southern
   part of the shock wave. This is further supported by the following analyses.

We also compare the shock heights measured from the MK4 observation and
that deduced from the type II emission using the Saito density model in Figure 4(a). The
shock heights are shown as bars whose length is due to uncertainties
of the density associated with the type II band width.
To measure the \textbf{heliocentric} distances of the northern and southern parts of the CME shock
wave, we draw two lines starting from the \textbf{source of the eruption passing through the
disturbed northern and southern streamers, respectively, and determine
the heliocentric distances of the intersection points of both lines with
the shock wave.} The two lines are plotted as solid and dashed lines in the upper
panels of Figure 3, and the obtained distances of the shock wave are shown
in Figure 4(a) as pluses and asterisks. We note  that
there are uncertainties in measuring these distances due to the quality
of the MK4 data. However, the large difference between the
distances along the two directions is significant. Therefore, we conclude that
the shock distances along the dashed line agree with the radio
shock heights. In other words, the radio source is likely located at
the southern part of the shock, rather than the northern part. This
strongly favors the \textbf{interaction of the shock wave with the
southern streamer} being the cause of the spectral bump.

To confirm the applicability of the Saito model to this event, in Figure 4(b) we
plot the radial electron density profiles deduced from the MK4 polarized
brightness (pB) data measured before the eruption at 22:30:06 UT. The
profiles are along 3 position angles (PAs) at 230, 240, and 270 degrees.
We use the standard pB inversion package from the SolarSoftWare (SSW). The Saito density
model is also shown as the solid line. Below 1.6 R$_\odot$,  all three sets of the pB density
profiles are considerably close to the Saito model. This indicates the applicability
of this model in a range of PAs, therefore the above conclusion should not be
affected by the significant non-radial propagation of the shock wave.

We now compare the estimated time of the shock transit across
the streamer with the bump duration. If the radio bumps
were caused by the shock passing through the southern streamer, it
had to be the shock flank, not the nose.  Assuming a propagation distance
of the shock inside the streamer being 0.5 R$_\odot$ - 1 R$_\odot$
(reasonable values for width of a streamer), and a shock speed of 800 km s$^{-1}$,
 then the transit will take 7.2 to 14.5 minutes. This is also
consistent with the 10-minute duration of the spectral bump.

In summary, we believe that the bumped type II burst in this event
is caused by electrons accelerated at the southern flank
 of the  CME-driven shock, and the bump is due to the
shock propagating through the dense streamer. Figure 5 is a
cartoon illustrating our understanding of this event. The outward
propagating CME-driven shock is denoted by the red solid-dashed
curve and the radio source by thick segment.
We can see that the streamer transit of the radio source
took place between 22:45 UT to 22:55 UT. This transit accounted
for the spectral bump reported above. We can also see that the
radio-emitting part of the shock is likely
quasi-perpendicular. We will return to this point in the
discussion section.

Last, we have pointed out that the first signature of the interaction
of the shock flank with the southern streamer seemed to occur
$\sim$ 4 minutes before the start of the spectral bump. This can be understood if
the radio source region (and therefore the associated energetic electron acceleration region)
is not at the edge of the shock flank, which entered the streamer first.

\subsection{Other radio features observed during the spectral bump}

In this subsection, we report other radio features observed in the
dynamic spectrum of this event. Since the radio data available
provide no spatial information of radio signals, these features do
not necessarily have causal relationship with the spectral bump.
We include them here because they occurred temporally close to the
spectral bump.

From the dynamic spectrum shown in Figure 2, we observe diffusive
type IV radio emission with frequencies above 70 MHz starting
around 22:42 UT, at which time the H branch of the type II burst
became intermittent and split while the F branch almost
disappeared. It seems that this type IV burst lasted significantly
longer than the duration of the type II bump. In addition, two
episodes of type III-like bursts were present at 22:48 UT and
22:49 UT with frequencies starting at $>$180 MHz. It is not possible
to determine the drifting direction of these type III-like bursts.
If they had positive drifts, i.e., they were reversal (RS) type
III-like, it would imply the existence of precipitating electrons
(from shock region). On the other hand, there was a strong type
III burst, starting at about 22:47 UT, right below the bump in the
dynamic spectrum. This lower type III burst is possibly due to
shock-accelerated electrons that are released into open field
lines during the shock-streamer interaction process. Motions of
energetic electrons accounting for the above-mentioned type IIIs
and type IV emissions have been plotted in the cartoon shown in
Figure 5.

\section{Conclusions and discussion}

In this paper we explore the possibility of using particular
features (spectral bump in this study) of type II radio bursts to
infer the underlying electron acceleration sites. In the event,
we have identified clear frequency bump in the type II dynamic
spectrum, which was interpreted as a result of the CME shock
entering nearby dense streamer structure. We propose that the
type II radio burst included here was closely
associated with CME-driven shock-streamer interaction. We also
documented a few possibly-relevant radio signatures including
emissions of type IV bursts above and type III-like bursts above
and below the type II spectral bump.

As the radio emitting region carried by the shock approaches the
dense streamer structure, both the F and H bands can get absorbed
or reflected depending on the exact density gradient along the
observing line of sight. This may explain the intermittency or
disappearance of radio signals before or at the early stage of the
type II bump. The streamer region is denser and slower than the
surrounding solar wind plasmas. A shock propagated into a streamer
can therefore be strengthened. This will affect the electron
acceleration process and the consequent radio emission intensity.

The observed type IV burst may be excited by electrons that are
accelerated and released by the shock into the confining streamer
structure during the interaction. The type III bursts observed
above and below the bump may also be produced by shock-accelerated
electrons, which escaped from the shock propagating anti-sunwards or
sunwards along open or large-scale closed field lines during the
shock-streamer interaction. Further studies on the effect of the
shock-streamer interaction on electron dynamics and radio emission
are required for a better understanding of the radio features presented here.

One implication of this study is that the CME-driven shock flank are important
sources of type II bursts. This is consistent with previous
studies (e.g., Reiner et al., 2003; Cho et al., 2011). At the
flank, the initial lateral expansion of the ejecta is very fast
and can drive a shock there. Since field lines near the surface of
the Sun are largely radial, the shock geometry at the shock flank is
likely quasi-perpendicular. Electron acceleration at a
quasi-perpendicular shock has been the subject of previous
theoretical and simulation works (Wu, 1984; Lee et al., 1996; Zank
et al., 1996; Guo {\&} Giacalone, 2010), as well
as some observational analyses (Holman {\&} Pesses, 1983; Bale et
al., 1999). It was suggested that a quasi-perpendicular shock is
an effective electron accelerator. However, it should be noted
that coronagraph images, upon which the study is based, are
two-dimensional projection of the complex three-dimensional
eruptive processes, structures that are not relevant to our study
may appear as relevant. And it is generally not possible to
determine the exact coronal shock geometry with available data sets.

The study presented here provides an indirect way of inferring the
acceleration site of electrons along the shock surface. This can
be important to understanding the effect of shock geometry on the
electron acceleration process. In the future, we plan to extend our studies to
examine more events with spectral bumps and understand their physical connection
to the CME (shock)-streamer interactions.

\acknowledgements We are grateful to the STEREO, SOHO/LASCO,
MLSO/MK4, Wind/Waves, and LEAR teams for making their data
available online. We thank Dr. Stephen White and Dr. Bill Erickson
for providing the BIRS data. This work was supported by 973 program 2012CB825601,
NNSFC grants 40825014, 40890162, 41028004, and the Specialized Research
Fund for State Key Laboratory of Space Weather in China. H. Q.
Song was also supported by NNSFC 41104113. GL's work at UAHuntsivlle was
supported by NSF CAREER: ATM-0847719 and NSF SHINE: AGS-0962658. GL would also like
to thank YC and his group for their hospitality during his visit at SDU at Weihai.

\newpage
\begin{figure}
% \epsscale{1.}
 \includegraphics[width=0.5\textwidth]{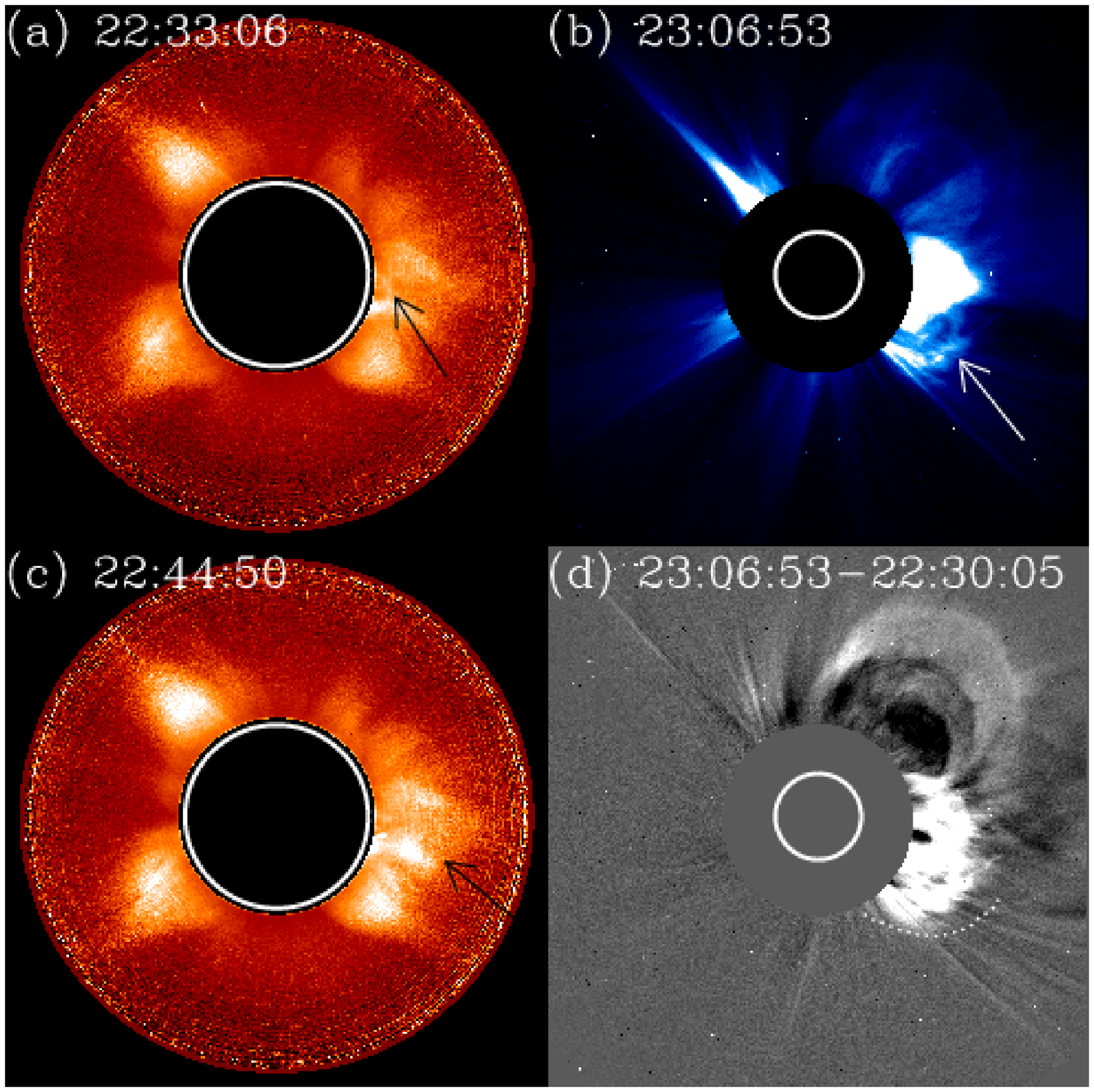}
\caption{Coronal images for the 2003 November 1 event observed by
MK4 and LASCO C2 coronagraphs (a - c). The arrows denote the location of the CME fronts. Panel
(d) is the difference image obtained from the LASCO C2
observations at 23:06:53 UT and 22:30:05 UT. Dotted curve in (d) plots
the diffusive shock ahead of the bright CME ejecta on the day. The CME
erupts from within a multi-streamer system with a bright arc-like
front, pushing aside the surrounding streamers during its rapid
expansion and propagation. (Animations of this figure are available in the online journal.)} \label{Fig:fig1}
\end{figure}

\begin{figure}
% \epsscale{1.}
 \includegraphics[width=0.8\textwidth]{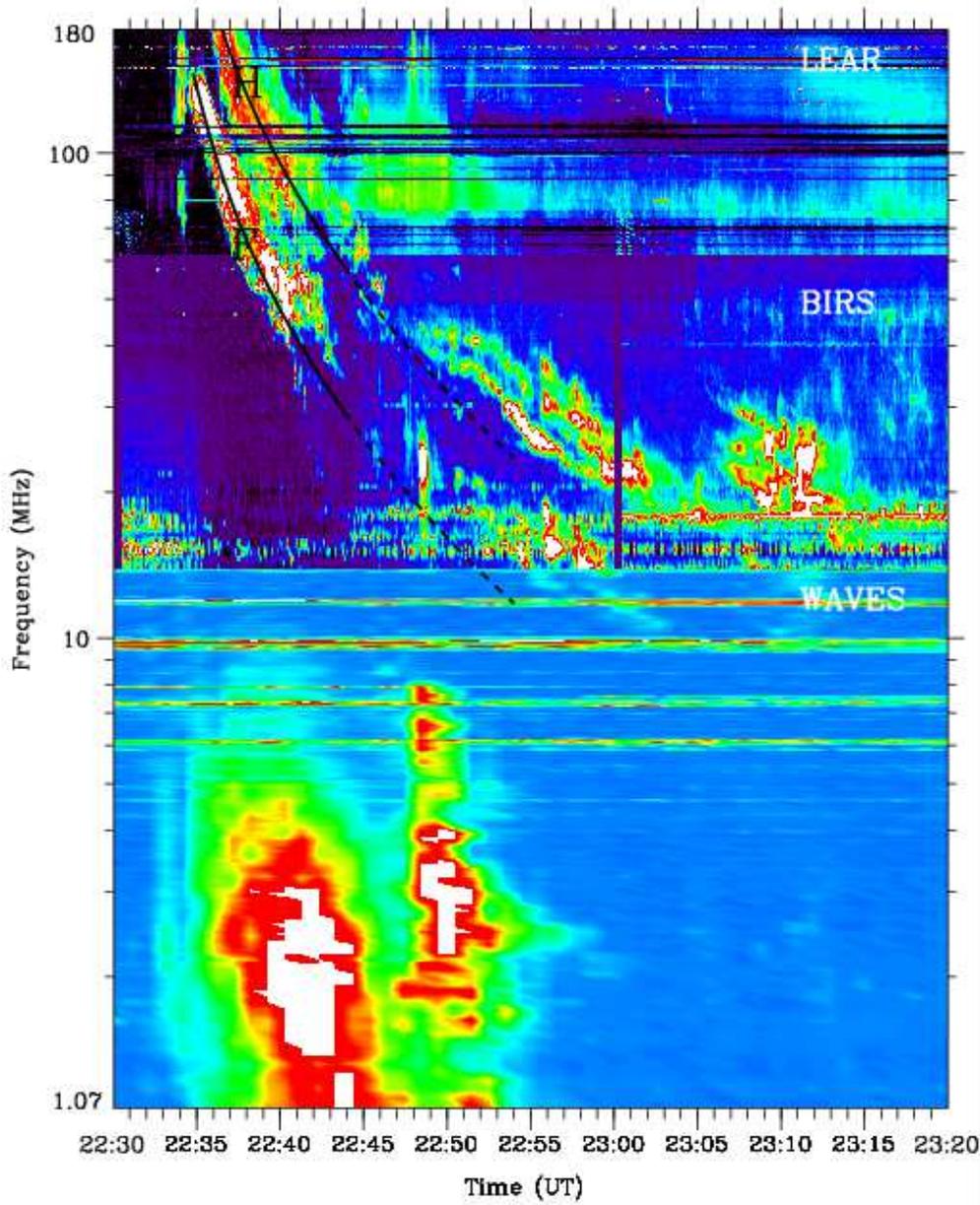}
\caption{The radio dynamic spectrum on 2003 November 1 from 22:30
UT to 23:20 UT, given by the composition of the spectra obtained
by WAVES (1.07-13.8 MHz), BIRS (13.8 - 62.5 MHz), and LEAR (62.5 -
180 MHz). F and H denote the fundamental and harmonic
branches of the burst. The two solid-dashed lines are fittings to the radio
spectrum before 22:44 UT using the one-fold Saito model and a shock
speed of 900 km s$^{-1}$.  } \label{Fig:fig2}
\end{figure}

\begin{figure}
% \epsscale{.8}
 \includegraphics[width=0.8\textwidth]{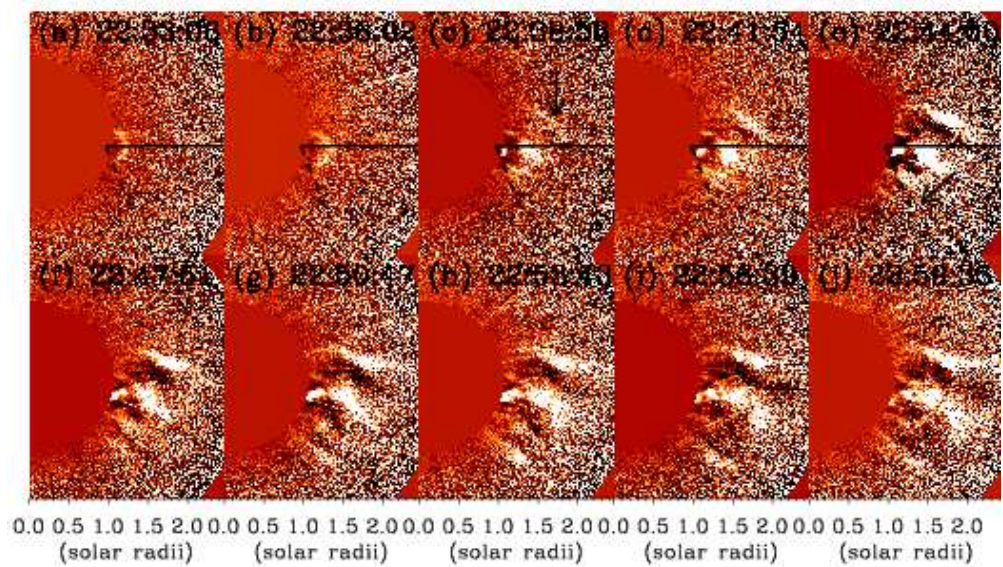}
\caption{Ten MK4 images for the 2003 November 1 event from 22:33:06 UT to 22:59:35 UT with 3 minutes
apart. All images have subtracted the pre-eruption corona image at
22:30:06 UT.  The solid and dashed lines in the upper panels are
drawn to measure the heights of the expanding CME fronts. } \label{Fig:fig3}
\end{figure}

\begin{figure}
% \epsscale{.8}
 \includegraphics[width=0.8\textwidth]{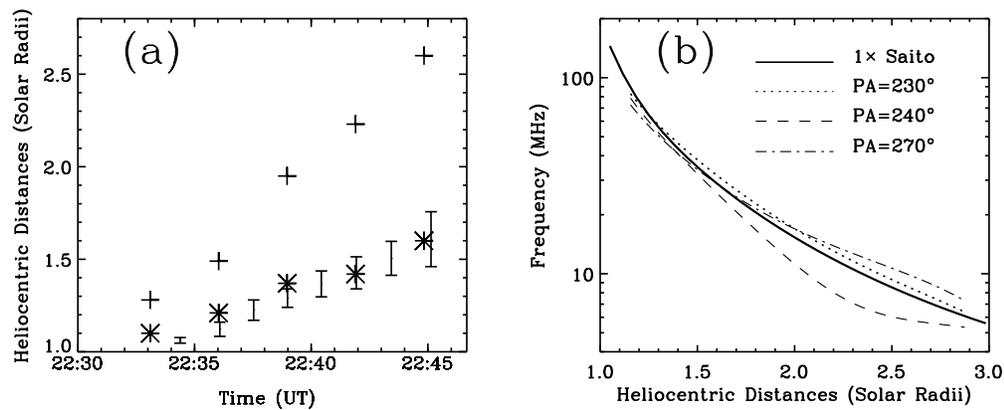}
\caption{(a) Heliocentric distances of the CME front measured
along the solid (pluses) and dashed (asterisks) lines plotted in
the upper panels of Figure 3. The bars are given by the shock
distances determined using the one-fold Saito density model (see
Figure 2). (b) The radial electron density profiles deduced from
the pre-eruption MK4 polarized brightness (pB) data at 22:30:06 UT,
along 3 PAs at 230, 240, and 270 degrees. The one-fold Saito
density model is also shown as the solid line.} \label{Fig:fig4}
\end{figure}

\begin{figure}
% \epsscale{1.}
 \includegraphics[width=0.5\textwidth]{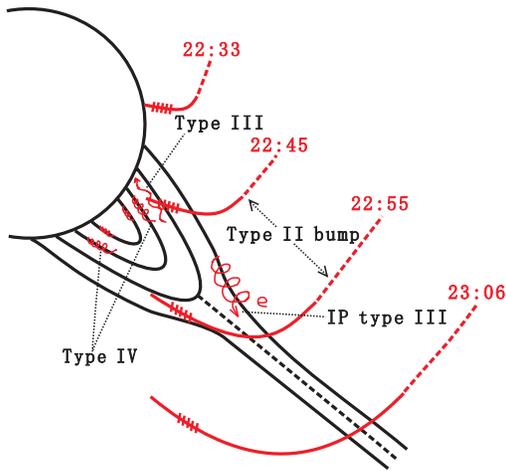}
\caption{A cartoon describing the magnetic topology of the
streamer structure, the outward propagation of the coronal shock wave,
and the location where the type II spectral bump took place. The
dashed line above the streamer cusp denotes the heliospheric
current sheet. The estimated type II source regions are shown as
the thick segments. The energetic electrons confined inside
the closed streamer arcades, and those flowing sunwards and
anti-sunwards are indicated. These electrons are related to type
IV, possible RS-III, and low-frequency type III bursts.}
\label{Fig:fig5}
\end{figure}

%\end{CJK*}

\begin{thebibliography}{}
\bibitem[Bale et al. 1999]{bale1999} Bale, S. D., Reiner, M. J., Bougeret, J. L.,
Kaiser, M. L., Krucker, S., Larson, D. E., Lin, R. P. 1999, GRL, 26, 1573
\bibitem[Bemporad et al. 2008]{bemporad2008}
Bemporad, A., Poletto, G., Landini, F., \& Romoli, M. 2008, Ann. Geophys., 26, 3017
\bibitem[Bemporad et al. 2010]{bemporad10}
Bemporad, A., Soenen, A., Jacobs, C., Landini, F., \&  Poedts, S. 2010, ApJ, 718, 251
\bibitem[Bougeret et al. 1995]{bougeret95}
Bougeret, J. L., Kaiser, M. L., Kellogg, P. J., Manning, R., Goetz, K., Monson, S. J.,
Monge, N., Friel, L., Meetre, C. A., \& Perche, C. 1995, Space Sci. Rev., 71, 231
%\bibitem[Bougeret et al. 1998]{bougeret98}
%Bougeret, J. L., Zarka, P., Caroubalos, C., Karlicky, M., Leblac, Y., Maroulis, D.,
% Hillaris, A., Moussas, X., Alissandrakis, C. E., Dumas, G., Perche, C. 1998, GRL, 25, 4103
\bibitem[Brueckner et al. 1995]{brueckner95}
Brueckner, G. E., Howard, R. A., Koomen, M. J., Korendyke, C. M., Michels, D. J.,
Moses, J. D., Socker, D. G., Dere, K. P., Lamy, P. L., \& Llebaria, A. et al. 1995,
Solar Phys., 162, 357.
%\bibitem[Cane et al. 1981]{cane1981}
%Cane, H. V., Stone, R.G., Fainberg, J., Steinberg, J. L., Hoang, S., Stewart, R. T.,
%1981, GRL, 8, 1285
%\bibitem[Cane \& Stone 1984]{cane1984}
%Cane, H. V., \& Stone, R. G. 1984, ApJ, 282, 339
\bibitem[Cane \& Erickson 2005]{cane05}
Cane, H. V., \& Erickson, W. C. 2005, ApJ, 623, 1180
\bibitem[Chen et al. 2001]{chen2001}Chen, Y., {\&} Hu, Y. Q. 2001,
Sol. Phys., 199, 371
\bibitem[Chen et al. 2010]{chen2010}Chen, Y., Song, H. Q., Li, B.,
Xia, L. D., Wu, Z., Fu, H., {\&} Li, X. 2010, ApJ, 714, 644
\bibitem[Cho et al. 2007]{cho2007}
Cho, K. S., Lee, J., Moon, Y. J., Dryer, M., Bong, S. C.,
Kim, Y. H., \& Park, Y. D. 2007, A\&A, 461, 1121
\bibitem[Cho et al. 2008]{cho2008}
Cho, K. S., Bong, S. C., Kim, Y. H., Moon, Y. J., Dryer, M., Shanmugaraju, A.,
Lee, J., \& Park, Y. D. 2008, A\&A, 491, 873
\bibitem[Cho et al. 2011]{cho2011}
Cho, K. S., Bong, S. C., Moon, Y. J., Shanmugaraju, A.,
Kwon, R. Y., \& Park, Y. D. 2011, A\&A, 530, 16
\bibitem[Cliver et al. 1999]{cliver1999}
Cliver, E. W., Webb, D. F., \& Howard, R. A. 1999, Sol. Phys, 187,89
\bibitem[Dulk 1985]{dulk85}
Dulk, G. A. 1985, ARA\&A, 23, 169
\bibitem[Elmore et al. 2003]{elmore2003}
Elmore, D. F., Burkepile, J. T., Darnell, J. A.,
Lecinski, A. R., \& Stanger, A. L. 2003, Proc. SPIE, 4843
\bibitem[Erickson 1997]{erickson1997}
Erickson, W. C. 1997, Publication Astronomical Society of
Austraila, 14, 3, 278-282
\bibitem[Feng et al. 2011]{feng2011}
Feng, S. W., Chen, Y., Li, B., Song, H. Q., Kong, X. L.,
Xia, L. D., \& Feng, X. S. 2011, Sol. Phys., 272, 119
\bibitem[Gergely 1983]{gergely1983}
Gergely, T. E., Kundu, M. R., Hildner, E. 1983, Sol. Phys., 268, 403
\bibitem[Ginzburg \& Zhelezniakov 1958]{ginzburg1958}
Ginzburg, V. L., \& Zhelezniakov, V. V. 1958, SvA, 2, 653
\bibitem[Guo \& Giacalone 2010]{guo2010}
Guo, F., \& Giacalone, J. 2010, ApJ, 715, 406
\bibitem[Habbal et al. 1997]{habbal1997}
Habbal, S. R., Woo, R., Fineschi, S., O'Neal, R., Kohl, J.,
Noci, G., \& Korendyke, C. 1997, ApJ, 489, 103
\bibitem[Holman \& Pesses 1983]{holman1983}
Holman, G. D., \& Pesses, M. E. 1983, ApJ, 267, 837
\bibitem[Kennewell \& Steward, 2003]{kennewell2003}
Kennewell, J. \& Steward, G. 2003, Solar Radio Spectrograph [SRS] Data Viewer
[Srsdisplay] (Sydney: IPS Radio and Space Serv.),
\url{ftp://ftp.ips.gov.au/wdc-data/spec/doc/Other\%20Document/Srsdispl.doc}
\bibitem[Kong et al. 2011 ]{kong2011}
Kong, X. L., Chen, Y., Feng, S. W., Song, H. Q., Li, G.,
Guo, F., Jiao, F. R. 2012, in press
\bibitem[Lee et al. 1996] {lee1996}
Lee, M., Shapiro, V. D., \& Sagdeev, R. Z. 1996, J. Geophys. Res., 101, 4777
\bibitem[Li \& Li 2006]{li2006}
Li, B., \& Li, X. 2006, A\&A, 456, 359
\bibitem[Liu et al. 2009] {liu09}
Liu, Y., Luhmann, J. G., Bale, S. D., \& Lin, R.P. 2009, ApJ, 691, L151
\bibitem[Magdaleni\' c et al. 2008]{magdalenic2008}
Magdaleni\' c, J., Vr\v snak, B., Pohjolainen, S., Temmer, M.,
Aurass, H., Lehtinen, N. J. 2008, Sol. Phys., 253, 305
\bibitem[Maia et al. 2000]{maia2000}
Maia, D., Pick, M., Vourlidas, A., Howard, R. 2000, ApJ, 528, 49
\bibitem[Nelson {\&} Melrose 1985]{nelson1985}
Nelson, G. J., {\&} Melrose, D. B. 1985, Solar Radiophysics: Studies of
Emission from the Sun at Metre Wavelengths, ed. D. J. McLean \& N. R. Labrum (Cambridge: Cambridge Univ. Press), 333
\bibitem[Nindos et al. 2011]{nindos2011} Nindos, A., Alissandrakis, C.E.,
Hillaris, A., Perka-Papadema, P. 2011, A\&A, 531, 31
\bibitem[Oh et al. 2007]{oh2007}
Oh, S. Y., Yi, Y., Kim, Y. H. 2007, Sol. Phys., 245, 391
\bibitem[Parenti et al. 2000]{parenti2000}
Parenti, S., Bromage, B. J. I., Poletto, G., Noci, G.,
Raymond, J. C., \& Bromage, G. E. 2000, A\&A, 363, 800
\bibitem[Pick 1999]{pick1999}
Pick, M., 1999,  in Proc. Nobeyama Symp., ed. T. Bastion, N.
Gopalswamy, \& K. Shibasaki (NRO Rep. 479; Kiysato: Nobeyama Radio Obs.), 187
\bibitem[Pick \& Vilmer 2008]{pick2008}
Pick, M., \& Vilmer, N. 2008, A\&ARv, 16, 1
\bibitem[Pohjolainen 2008]{pohjolainen2008}
Pohjolainen, S. 2008, A\&A, 483, 297
\bibitem[Reiner et al. 2003]{reiner2003}
Reiner, M. J., Vourlidas, A., \& Cyr, O. C. St. et al. 2003, ApJ, 590, 533

%\bibitem[Robinson \& Sheridan 1982]{robbinson1982} Robinson, R. D., \& Sheridan, K. V. 1982,
% Proc. Astron. Soc. Aust. 4, 392
\bibitem[Saito 1970]{saito1970}Saito, K., 1970, Ann. Tokyo Astr. Obs., 12, 53

\bibitem[Sheeley et al. 2000]{sheeley2000}
Sheeley, N. R., Hakala, W. N., Wang, Y. M. 2000, J. Geophys. Res. 105, 5081

\bibitem[Smerd et al. 1974]{smerd1974}
Smerd, S. F., Sheridan, K. V., \& Stewart, R. T. 1974,
in IAU Symp. 57, ed. G. A. Newkirk, 389
\bibitem[Strachan et al. 2002]{strachan2002}
Strachan, L., Suleiman, R., Panasyuk, A. V.,
Biesecker, D. A., \& Kohl, J. L. 2002, ApJ, 571, 1008
\bibitem[Subramanian et al. 1999] {subramanian1999}
Subramanian, P., Dere, K. P., Rich, N. B., \& Howard, R. A. 1999,
J. Geophys. Res., 104, 321
\bibitem[Vourlidas et al. 2003]{vourlidas2003}
Vourlidas, A., Wu, S. T., Wang, A. H., Subramanian, P., \& Howard,
R. A. 2003, ApJ, 598, 1392
\bibitem[Vr\v snak {\&} Cliver]{vrsnak2001} Vr\v snak, B., Aurass, H.,
Magdalenic, J., Gopalswamy, N. 2001, A\&A, 377, 321
\bibitem[Vr\v snak {\&} Cliver]{vrsnak2008} Vr\v snak, B.,
Cliver, E. W. 2008, Sol. Phys., 253, 215
\bibitem[Wang et al. 2007] {wang2007}
Wang, Y. M., Sheeley, N. R. Jr., \& Rich, N. B. 2007, ApJ, 658, 1340
\bibitem[Wild {\&} McCready 1950]{wild1950} Wild, J. P.,
{\&} McCready, L. L. 1950, Aust. J. Phys. 3, 387
\bibitem[Wild et al. 1954]{wild1954} Wild, J. P.,
Murray, J. D., Rowe, W. C. 1954, Aust. J. Phys. 7, 439
\bibitem[Wu 1984]{wu1984} Wu, C. S. 1984, J. Geophys. Res., 89(A10), 8857
\bibitem[Yashiro 2004]{yashiro2004}
 Yashiro, S., Gopalswamy, N., Michalek, G., St. Cyr, O. C., Plunkett, S. P.,
 Rich, N. B., Howard, R. A., J. Geophys. Res., 2004, 109(A7), 07105
\bibitem[Zank et al. 1996]{zank1996}
Zank, G. P., Pauls, H. L., Cairns, I. H., {\&} Webb, G. M. 1996,
 J. Geophys. Res., 101, 457

\end{thebibliography}
\end{document}